\documentclass[aps,prb,twocolumn,floatfix,superscriptaddress]{revtex4-1}
\usepackage[utf8]{inputenc}
\usepackage{graphicx}
\usepackage{epstopdf}
\usepackage{float}
\usepackage{color}
\usepackage{xcolor,colortbl}
\usepackage{amsmath}
\usepackage[caption=false,subrefformat=parens]{subfig}
\usepackage{comment}
\usepackage{xr}

\newcommand{\bea}{\begin{eqnarray*}}
\newcommand{\eea}{\end{eqnarray*}}
\newcommand{\bne}{\begin{equation*}}
\newcommand{\ede}{\end{equation*}}

\newcommand{\bnen}{\begin{equation}}
\newcommand{\eden}{\end{equation}}
\newcommand{\bean}{\begin{eqnarray}}
\newcommand{\eean}{\end{eqnarray}}
\newcommand{\bnsn}{\begin{subequations}}
\newcommand{\edsn}{\end{subequations}}

\begin{document}
\title{\emph{Ab initio} spin-strain coupling parameters of divacancy qubits in silicon carbide}
\author{P\'eter Udvarhelyi}
\affiliation{Department of Biological Physics, Lor\'and E\"otv\"os University, 
P\'azm\'any P\'eter s\'et\'any 1/A, H-1117 Budapest, Hungary}
\affiliation{Wigner Research Centre for Physics, Hungarian Academy of Sciences, P.O. Box 49, H-1525 Budapest, Hungary}
\author{Adam Gali}
\affiliation{Wigner Research Centre for Physics, Hungarian Academy of Sciences, P.O. Box 49, H-1525 Budapest, Hungary}
\affiliation{Department of Atomic Physics, Budapest University of Technology and Economics, Budafoki \'ut 8., H-1111 Budapest, Hungary}
\date{\today}

\begin{abstract}
Cubic silicon carbide is an excellent platform for integration of defect qubits into established wafer scale device architectures for quantum information and sensing applications, where divacancy qubit, that is similar to the negatively charged nitrogen-vacancy (NV) center in diamond, has favorable coherence properties. We demonstrate by means of density functional theory calculations that divacancy in 3C SiC has superior spin-stress coupling parameters and stress sensitivity for nanoscale, quantum enhanced photonic, optoelectronic and optomechanical devices.
\end{abstract}

\maketitle

\section{Introduction}
\begin{figure}
\centering
\includegraphics[scale=0.20]{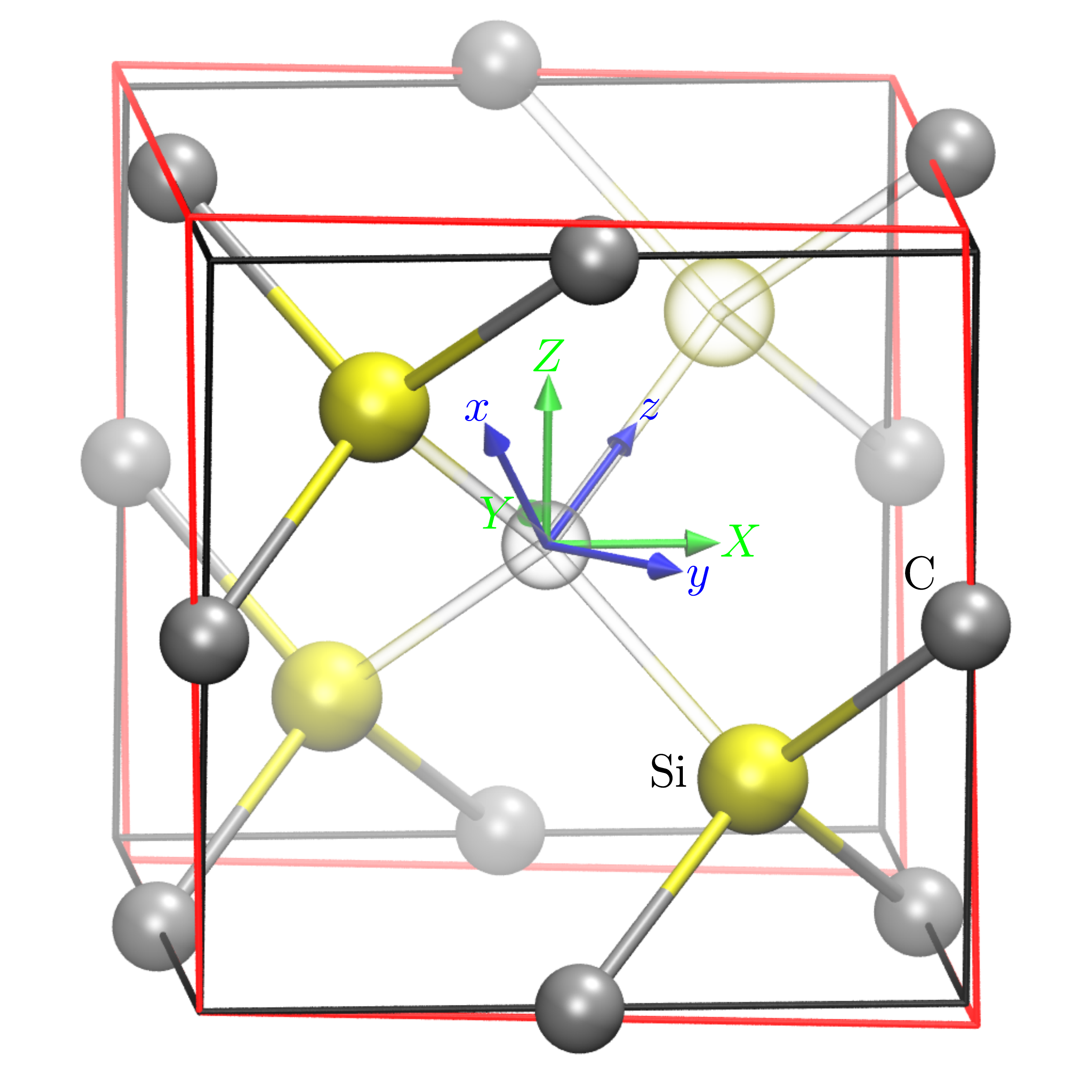}
\caption{Divacancy in silicon carbide (cubic Bravais cell shown in black). $\{XYZ\}$ shows the crystal reference frame and $\{xyz\}$ defines the local reference frame of the center. Deformed cell is visualized in red for $\varepsilon_{xx}=0.1$ strain.}
\label{fig:divacancy}
\end{figure}

Silicon carbide (SiC) is an emerging host for qubit defects~\cite{Gali2011, Weber2010, Koehl2011, Baranov2011, Riedel2012, Kraus2013}. The main advantage of SiC as a host material is its industrial scale availability, high quality single crystal growth in substrate scale and epitaxial thin layer growth on silicon wafer~\cite{Zorman}. Furthermore, advanced microfabrication techniques are already available for potential integration of spin qubit sensors into semiconductor devices. In particular, divacancy spins exhibit optical addressability and long coherence time in the most common polytypes of SiC, cubic 3C, and hexagonal 4H and 6H~(Ref.~\onlinecite{Falk2013}). In this paper we focus on the 3C polytype and its neutral divacancy defect consisting of neighboring silicon and carbon vacancy (see Fig.~\ref{fig:divacancy}). The divacancy qubit in 3C SiC is especially interesting because nanoelectromechanical-sensors (NEMS) can be produced from thin films of 3C SiC that was grown on silicon wafers~\cite{Yang2001}, and it has been shown~\cite{Calusine2014} that divacancy qubits can be engineered into these 3C SiC thin films. The fingerprints of 3C divacancy are the Ky5 electron paramagnetic resonance (EPR) center~\cite{Bratus} with $S=1$ spin and L3 optically detected magnetic resonance (ODMR) center~\cite{Son} with near infrared (NIR) photoluminescence line at 1.12~eV. The defect has remarkable Hahn echo coherence time of 0.9~ms in 3C SiC (Ref. \onlinecite{Christle2017}), similar to the 1.2~ms coherence time of divacancies in 4H SiC (Ref.~\onlinecite{Christle2014}), observed in natural isotope abundant samples at 20~K. We conclude that the divacancy NIR color center has similar spin and optical properties to the negatively charged nitrogen-vacancy (NV) center in diamond~\cite{Gali2011, Christle2017}, even surpassing its coherence time of 0.6~ms~\cite{Stanwix, Christle2017}. These 3C divacancy qubits in NEMS can be harnessed to measure strain at the nanoscale.  

Although NV center is presently the most studied nanoscale strain sensor~\cite{TeissierPRL, Barfuss2015, Ovartchaiyapong2014, BarsonNL, MacQuarrie_dynamical, MacQuarrie_NatComm2017, MacQuarrie:15, Golter_PRL2016, GolterPRX, Meesala}, the diamond host suffers from difficulties in crystal growth and fabrication at large scale. Thus, finding alternative defect qubit nanoscale sensor in technologically mature materials, such as SiC, with similar or superior sensitivities is of high importance. The straightforward production of 3C SiC thin films hosting divacancy qubits with favorable coherence properties makes 3C divacancy a very attractive potential candidate in realizing strain sensors at the nanoscale. However, the strengths of spin-strain couplings for the divacancy qubit in 3C SiC have not been determined so far that are critical parameters in the sensitivity of future pressure and electromechanical sensors. We note that, parallel to our study, qubits controlled by alternating pressure and electric fields have been demonstrated as 4H SiC divacancies of which phenomenon should rely on considerable spin-strain coupling in these qubits~\cite{Whiteley2018}, that further strengthens the need of studying the spin-strain coupling parameters of divacancies in SiC.

In this paper, we calculate the spin-strain coupling parameters of divacancy qubit in 3C SiC by means of first principles calculations, and estimate the stress sensitivity with taking realistic key parameters of the divacancy qubit and the host SiC. We show that divacancy qubits in SiC have generally larger spin-stress coupling parameters than that of NV center in diamond, where the stiffness of SiC gives rise to this phenomena. As a consequence, the sensitivity of SiC divacancy qubits can be harnessed to realize nanoscale, quantum enhanced photonic, optoelectronic and optomechanical devices on a platform that is compatible with semiconductor technology and electronics.

\section{Methods}
\subsection{Spin-strain Hamiltonian}
Recent works have been carried out to describe the spin-strain coupling parameters of NV center in diamond~\cite{TeissierPRL, Barfuss2015, Ovartchaiyapong2014, BarsonNL, MacQuarrie_dynamical, MacQuarrie_NatComm2017, MacQuarrie:15, Golter_PRL2016, GolterPRX, Meesala}. Since NV center and divacancy in 3C SiC share the same symmetry the spin-strain Hamiltonian developed for NV center in diamond can be directly applied to divacancy in 3C SiC. Very recently, we have advanced and completed the theory for the spin-strain Hamiltonian for the NV center~\cite{Udvarhelyi} which has the form
\begin{subequations}
\label{eq:spinstrain}
\bean
H_{\varepsilon} &=& H_{\varepsilon 0} + H_{\varepsilon 1}+H_{\varepsilon 2},
\\
H_{\varepsilon 0} /h&=&
[h_{41} (\varepsilon_{xx} + \varepsilon_{yy}) 
+
h_{43} \varepsilon_{zz}]S_z^2,
\label{eq:h0}
\\
\label{eq:h1}
H_{ \varepsilon 1}/h &=& \nonumber
\frac 1 2
\left[h_{26} \varepsilon_{zx}
- \frac 1 2 h_{25} (\varepsilon_{xx} - \varepsilon_{yy})
\right]
 \{ S_x,S_z\}
\\
&+&
\frac 1 2 
\left(
h_{26}  \varepsilon_{yz} 
+  h_{25} \varepsilon_{xy}
\right)
\{ S_y,S_z\},
\\
H_{\varepsilon 2} /h&=&  \frac 1 2 \left[
	 h_{16}  \varepsilon_{zx}
	- \frac 1 2 h_{15} (\varepsilon_{xx} - \varepsilon_{yy})
\right](S_y^2-S_x^2) \nonumber
\\
&+& \frac 1 2 (
 h_{16} \varepsilon_{yz} + h_{15} \varepsilon_{xy}
) \{S_x, S_y\},
\eean
\end{subequations}
where $\varepsilon_{ij} = (\partial u_i/\partial x_j +\partial
u_j/\partial x_i)/2$ are the Cartesian elements of the strain tensor and ${\bf u}({\bf r})$
is the displacement field. The spin-strain coupling parameters are labeled by $h$. 
The spin-stress Hamiltonian has the same symmetry-allowed form with coupling parameters labeled by $g$.

To calculate spin-stress coupling parameters
from spin-strain coupling parameters,
we use the stiffness tensor $C$ of bulk 3C SiC, with elements in the cubic reference frame: $C_{11}=390~\mathrm{GPa}$, $C_{12}=142~\mathrm{GPa}$, 
$C_{44}=256~\mathrm{GPa}$ (experimental data derived by Lambrecht {\it et al.}\cite{PhysRevB.44.3685} from measurements of Feldman {\it et al.}\cite{PhysRev.173.787}).
First, we transform the stiffness tensor to the defect frame;
we denote the resulting $6\times 6$ stiffness matrix 
in the Voigt notation as $C$. 
To convert the spin-strain Hamiltonian 
Eq.~\eqref{eq:spinstrain} 
to spin-stress Hamiltonian,
we express the strain components in Eq.~\eqref{eq:spinstrain}
using stress components via
$\varepsilon = C^{-1} \sigma$, 
where $\varepsilon = (\varepsilon_{xx},\varepsilon_{yy},\varepsilon_{zz},
2\varepsilon_{yz},2\varepsilon_{zx},2\varepsilon_{xy})$
and $\sigma = (\sigma_{xx},\sigma_{yy},\sigma_{zz},\sigma_{yz},\sigma_{zx},\sigma_{xy}) $ holds in Voigt notation.

\subsection{\emph{Ab initio} spin-strain coupling parameters}
We determined the 
spin-strain coupling parameters 
using density functional theory (DFT). 
We applied DFT for electronic structure calculation 
and geometry optimization,
using the PBE functional\cite{PBE} 
in the plane-wave-based 
Vienna Ab initio Simulation Package (VASP)\cite{VASP1,VASP2,VASP3,VASP4}. 
The core electrons were treated in the projector augmented-wave 
(PAW) formalism\cite{paw}. 
The calculations were performed with $600~\text{eV}$ 
plane wave cutoff energy. 
The model of the divacancy in bulk 3C SiC was constructed 
using a 512-atom simple cubic supercell within the 
$\Gamma$-point approximation. 
We use a negative sign convention for compressive strain.
The model of mechanical strain, described by the strain tensor $\varepsilon$, 
was the deformed cubic supercell with edge vectors obtained by transforming the
nondeformed edge vectors with the matrix 
$1+\varepsilon$ in the cubic reference frame,
and allow the atomic positions to relax.
For each strain configuration, 
the elements of the 
$3 \times 3$
zero-field splitting matrix $D$,
defining the ground-state spin Hamiltonian
via $H = \vec S^T \cdot  D \cdot \vec S$, 
were calculated using the VASP
implementation by Martijn Marsman with
the PAW formalism~\cite{Bodrog}. We calculated the deformed supercells at several points and we applied a linear regression to read out the coupling-strength parameters as explained in Ref.~\onlinecite{Udvarhelyi}.

In order to test the accuracy of our method, we studied the $hh$ divacancy qubit in 4H SiC, for which experimental data was available~\cite{PhysRevLett.112.187601}. Our results show good agreement with the observed spin-strain couplings (see Appendix~\ref{app:4H}).

\section{Results and discussion}

The \emph{ab initio} spin-strain and the derived spin-stress coupling parameters of the neutral divacancy in 3C SiC are summarized in Table~\ref{tab:DFT3C}. The $h_{25}$ and $h_{26}$ couplings are responsible for the flipping of the electron spin which are comparable to the other coupling parameters. This holds for the 
$hh$ divacancy in 4H SiC (see Appendix~\ref{app:4H}) which explains the recently observed electromechanical driving of these spins~\cite{Whiteley2018}. By comparing these results with our recent data on NV center~\cite{Udvarhelyi}, we realize that 3C divacancy exhibits slightly smaller spin-strain coupling parameters but greater spin-stress coupling parameter for most of the types of distortion 
than diamond NV does. This is caused by the smaller stiffness parameters of 3C SiC than those of diamond. Our results demonstrate that the mechanical properties of the host material can seriously affect the final response of the embedded qubits to external stress.
\begin{table}
\caption{
Spin-strain ($h$) and spin-stress ($g$) coupling parameters of divacancy in 3C SiC as obtained from density functional theory. Results are rounded to significant digits.}
\begin{ruledtabular}
\begin{tabular}{lll}
parameter & h (MHz/strain) & g (MHz/GPa)\\\hline
$h_{43}$, $g_{43}$ & $2530\pm30$ & $6.01\pm0.07$\\
$h_{41}$, $g_{41}$ & $-4700\pm200$ & $-8.1\pm0.3$\\
$h_{25}$, $g_{25}$ & $-900\pm100$ & $-0.7\pm0.3$\\
$h_{26}$, $g_{26}$ & $-1760\pm20$ & $-5.00\pm0.1$\\
$h_{15}$, $g_{15}$ & $3200\pm200$ & $7.1\pm0.5$\\
$h_{16}$, $g_{16}$ & $1320\pm50$ & $1.3\pm0.3$\\
\end{tabular}
\end{ruledtabular}
\label{tab:DFT3C}
\end{table}

\subsection{Stress sensitivity of 3C divacancy based on ODMR readout}

We discuss here how the spin-stress coupling in 3C divacancy can be harnessed in nanoscale sensing applications. The most common readout mechanism of the defect spins is the ODMR method. In this case, the shot noise-limited sensitivity for sensing magnetic fields, electric fields, temperature and strain in a Hahn echo measurement is generally written in the form~\cite{Pham}
\begin{equation}
\label{eq:sensitivity}
\eta=\frac{1}{4gC\sqrt{\beta T_{2}}}\text{,}
\end{equation}
where $g$ is the coupling parameter to spin, $T_{2}$ is the homogeneous spin coherence time, and $C$ is the fluorescence readout contrast. Here we approximated the measurement time and free precession time by $T_{2}$. The contrast $C$ is defined as
\begin{equation}
C=\frac{p_{0}-p_{1}}{p_{0}+p_{1}}
\end{equation}
with $p_{0}$ and $p_{1}$ detected photon counts in the bright and dark state, respectively. $\beta$ is the average fluorescence intensity that is approximated by $p_{0}$. The truly intrinsic parameter associated with the qubit in $\eta$ is the $g$ coupling parameter. The photon counts and $T_2$ time depend on the quality and shape of the host material and other experimental conditions. 

The off-resonant readout contrast of isolated 3C divacancy is about $C=7.5\%$, the saturation photon count rate is $26~\mathrm{kcts/s}$ and the spin coherence time is $T_{2}=0.9~\mathrm{ms}$ in nearly dopant-free crystal at 20~K temperature\cite{Christle2017}. We note that the quality of 3C SiC samples still did not reach the quality of 4H SiC samples because 4H SiC is employed in SiC semiconductor devices that has driven the improvement of specifically 4H polytype of SiC. We note that further reduction of the nitrogen donor concentration in 3C SiC to the typical values of high quality 4H SiC would certainly converge the coherence times of the divacancies in the two polytypes, and the ODMR contrast of divacancy in 3C SiC can be further improved by optimizing the excitation wavelength and lowering the background from other defects, similarly to the divacancy qubits in 4H SiC. Therefore, we assume that these high quality 3C SiC samples are in reach, and we estimated the sensitivity of 3C divacancy from 4H SiC data of $C=15\%$ and $T_{2}=1.2~\mathrm{ms}$ (Ref.~\onlinecite{Christle2014}). We assumed the same ODMR readout time for 3C divacancy as was reported for diamond NV center at about $350~\mathrm{ns}$ (Ref.~\onlinecite{Toyli2012}), from which one can estimate the total photon counts during single readout event. Finally, we have all the parameters that enter in Eq.~\eqref{eq:sensitivity}. We find that $\eta\sim10^{-5}\text{GPa Hz}^{-1/2}$ for 3C divacancy qubits.

\begin{figure}
\includegraphics[width=0.9\columnwidth]{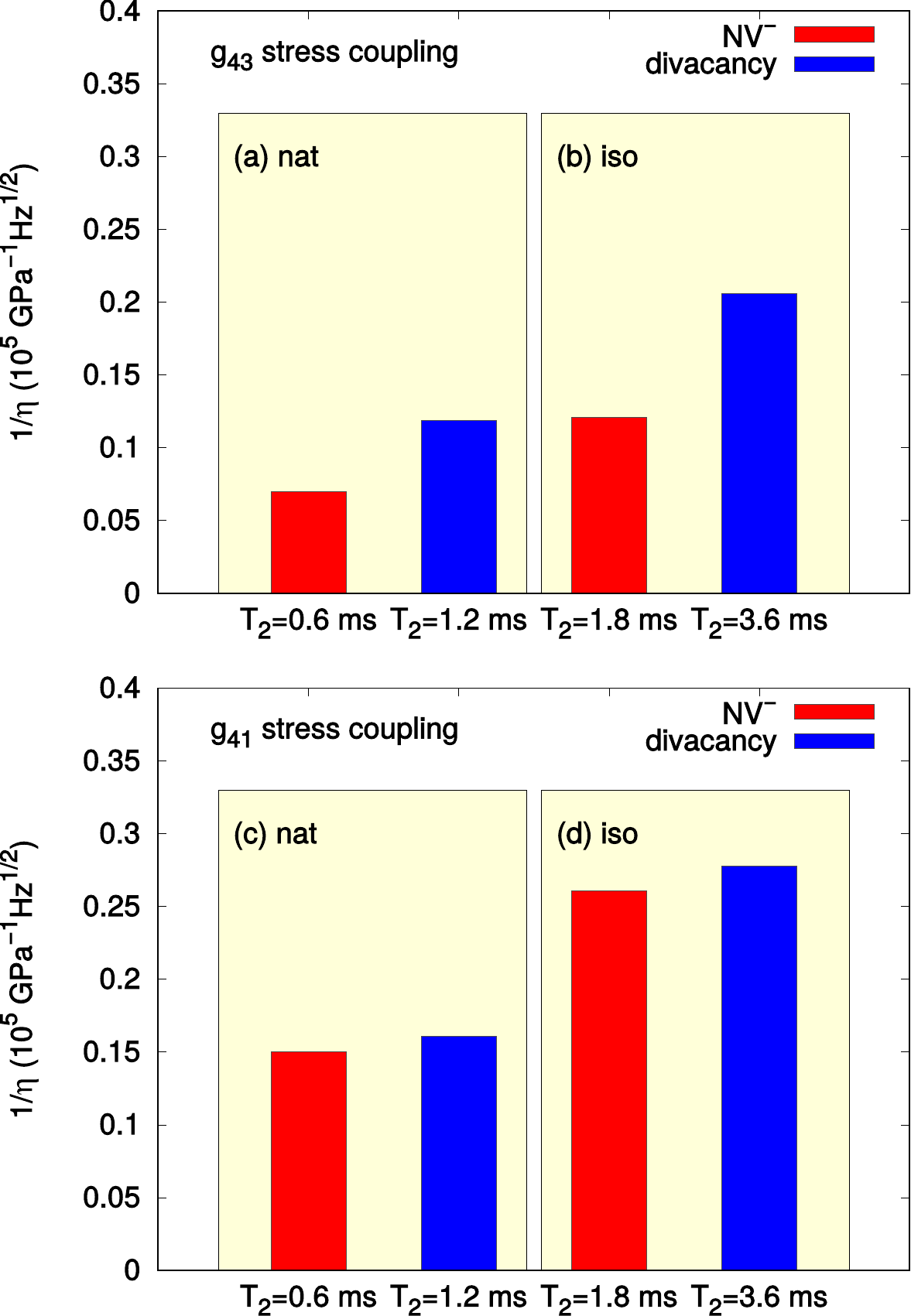}
\caption{Inverse stress sensitivity ($1/ \eta$) comparison of the negatively charged nitrogen-vacancy center in diamond (red column) and divacancy in silicon carbide (blue column). Estimated values for natural abundant crystals (nat) and isotopically purified samples (iso)  are shown in panels (a) and (c) and panels (b) and (d), respectively, where the corresponding $T_2$ times are depicted. Other parameters used in the calculations of ($1/ \eta$) are discussed in the text.}
\label{fig:comparison}
\end{figure}

We illustrate the results as blue columns in Fig.~\ref{fig:comparison}(a) and (c)
where we show the inverse of $\eta$ which implicates that the larger is the value
(height of the
column) the better is the sensitivity. In particular, we plot the inverse sensitivities
of $g_{43}$ and $g_{41}$ coupling parameters that corresponds to the pressure
along the symmetry axis and in the plane perpendicular to the symmetry axis,
respectively. We note that the former corresponds to the $c$-axis in the hexagonal
4H SiC lattice in Appendix~\ref{app:4H}. We find more sensitivity toward $g_{41}$
over $g_{43}$ but the order of magnitude is the same. 

We note that the $T_2$ time can be greatly improved by isotope engineering of the host material, i.e., by removing the nuclear spin noise as demonstrated for NV center in diamond~\cite{Balasubramanian2009}. They could increase the $T_2=0.6~\mathrm{ms}$ (Refs.~\onlinecite{Maze2008, Mizuochi2009}) up to $T_2=1.8~\mathrm{ms}$. Tight control of isotope engineering of 4H SiC was already demonstrated~\cite{Ivanov2014} that can be basically perpetuated for 3C SiC. We estimate similar improvements on 3C divacancies' $T_2$ time going from natural abundant to isotopically purified SiC samples that results in $T_2=3.6~\mathrm{ms}$. The corresponding results are shown as blue columns in Fig.~\ref{fig:comparison}(b) and (d) which show almost a factor of two improvement in the sensitivity.

We finally compare the sensitivities of 3C divacancy to that of NV diamond. For NV center in diamond, the off-resonant contrast is about $30\%$, the photon count rate is about $28~\mathrm{kcts/s}$ in mechanical resonator experimental setup, so we could estimate the sensitivity with these and the previously mentioned parameters. We find similar but smaller sensitivities for diamond NV center (red columns in Fig.~\ref{fig:comparison}) than for 3C divacancy. In particular, the sensitivity for $g_{43}$ coupling parameter of divacancy in isotopically purified 3C SiC samples is clearly superior over isotopically purified diamond NV center.

\section{Conclusion}
We have calculated the spin-strain coupling parameters for divacancy center in 3C SiC. In comparison to the most promising NV center in the field of nanoscale sensing, the intrinsic stress-spin coupling parameters of divacancy are superior. The actual sensitivity of the 3C divacancy depends on the quality of the 3C SiC crystal. We estimated that improvement on the quality of the 3C SiC crystal leads to favorable sensitivity of 3C divacancy nanosensors.  Non-optical spin readout techniques such as photocurrent detection of magnetic resonance (PDMR)\cite{Bourgeois2017} may further improve sensitivity for both centers by substituting the low photon collection efficiency with a high photocurrent efficiency. In particular, by realizing PDMR on 3C divacancy on Si substrate, an all-silicon based electronic chip sensor could be constructed for nanoscale measurement of pressure and electric fields. 

\section*{Acknowledgement}
We thank for the support of NKFIH within the Quantum Technology National Excellence Program  (Project No.~2017-1.2.1-NKP-2017-00001).

\appendix

\section{4H-SiC divacancy spin-strain coupling parameters}\label{app:4H}

DFT calculated spin-strain coupling parameters of $hh$ divacancy (PL1) in 4H-SiC are summarized in Table \ref{tab:DFT4H} where the defect was modeled in a 576-atom supercell with $\Gamma$-point sampling. The experimental ODMR shift for perpendicular strain was reported $(2-4)~\mathrm{GHz/strain}$ (Ref.~\onlinecite{PhysRevLett.112.187601}) corresponding to our calculated $h_{41}=5~\mathrm{GHz/strain}$ coupling parameter. The good agreement validates our DFT method for calculating divacancy's spin-strain coupling parameters in 3C-SiC too. For the stress conversion, we used the stiffness tensor elements of 4H SiC in the cubic reference frame
$C_{11}=507~\mathrm{GPa}$, $C_{12}=108~\mathrm{GPa}$, $C_{13}=52~\mathrm{GPa}$, $C_{33}=547~\mathrm{GPa}$, $C_{44}=159~\mathrm{GPa}$
 (Ref.~\onlinecite{Kamitani1997}).
\begin{table}[H]
\caption{
Spin-strain ($h$) and spin-stress ($g$) coupling-strength parameters of $hh$ divacancy in 4H-SiC
calculated from density functional theory. Results are rounded to significant digits.}
\begin{ruledtabular}
\begin{tabular}{lll}
parameter & h (MHz/strain) & g (MHz/GPa)\\\hline
$h_{43}$, $g_{43}$ & $3110\pm30$ & $7.33\pm0.06$ \\
$h_{41}$, $g_{41}$ & $-4940\pm60$ & $-8.65\pm0.09$ \\
$h_{25}$, $g_{25}$ & $1130\pm40$ & $2.8\pm0.1$ \\
$h_{26}$, $g_{26}$ & $-1580\pm30$ & $-5.0\pm0.1$ \\
$h_{15}$, $g_{15}$ & $7600\pm300$ & $18.9\pm0.6$  \\
$h_{16}$, $g_{16}$ & $1600\pm60$ & $5.0\pm0.2$  \\
\end{tabular}
\end{ruledtabular}
\label{tab:DFT4H}
\end{table}

%

\end{document}